# Development of A Modified Quasi-periodic Undulator for HLS

Yufeng Yang[1], Huihua Lu[1], Wan chen[1], Qika Jia[2], Shuchen Sun[1], Zhiqiang Li[1]

**1.** Institute of High Energy Physics, Chinese Academy of Sciences, 100049, Beijing, China

**2.** University of Science and Technology of China, 230029, Hefei, Anhui, China

**Abstract:** A modified quasi periodic undulator (QPU) is developed and to be installed at Hefei Light Source (HLS). Magnet dimensions optimization is applied. High harmonics contamination is eliminated from the fundamental emission effectively according to the field tests. The depression ratio of 3$^{rd}$ harmonic is increased by an order of magnitude than the current device with other harmonics well reduced simultaneously. The significance of the modification measure is verified practically. The design, measurement and commissioning of the device is described.

## 1. Introduction

The spectrum of a conventional undulator is made of a series of peaks harmonically related in which the energy of the harmonics occurs at a frequency precisely equal to odd number times that of the fundamental peak. The mixing of rational harmonics is not welcome because it leads to a degradation of the signal-to-noise-ratio for some user's experiments. However, the so-called monochromatic beam is always contaminated, to some extent, by the harmonics due to the limit of monochromator [1-2].

In recent years, effort has been made on quasi periodic undulator to depress the rational harmonic emission. The basic idea is by displacing some magnet bolcks in the Halbach undulators in an irrational way, rational harmonics is replaced by irrational harmonics[1-6]. One of the efficient scheme is ESRF type where some specially selected H-magnet, i.e., magnet horizontally magnetized, in a Halbach permanent magnet undulator are vertically retracted by a value $\delta$. In this way 3$^{rd}$ harmonic is reduced by an order of magnitude, as verified by a practical device [4].

A quasi periodic undulator is urged in the major reformation of HLS. Users hope the light beam can improve its monochromaticity as far as possible in which 3$^{rd}$ harmonic is assumed the leading contamination to the fundamental. This letter introduces the development of HLS QPU regarding to design, manufacturing and commissioning. To enhance harmonics depressions, It modifies ESRF type QPU by optimizing longitudinal dimensions of two magnet types which is proposed theoretically[7-8].

Magnetic tests were made to evaluate the undulator performance. Deduced spectrum is compared with the measured of ESRF QPU. The significance of magnet dimensions optimization on harmonics depression is inspected. 1$^{st}$ integrals in a good field region under tunable gaps is investigated.

## 2. Design

On the basis of ESRF type QPU, where an order of magnitude depression of high harmonic is presented, a modified scheme is advanced in theory to explore the possibilities of depressing harmonics more thoroughly [7-8]. Different from conventional pure permanent magnet undulators with uniform magnet sizes, the modified QPU takes magnets with respective longitudinal dimensions corresponding to different magnetization (Fig.1). Analytically, ESRF type QPU has two parameters $\delta$ and $\eta$ where $\eta$ is an irrational dimensionless number called the inter-lattice ratio. Modified QPU has another parameter $L_V/L_H$ besides those, where $L_V$ and $L_H$ are the longitudinal dimensions of magnets with vertical (V-magnet) and horizontal (H-magnet) magnetization. There is $L_V + L_H = \lambda_u/2$ where $\lambda_u$ is the period length.

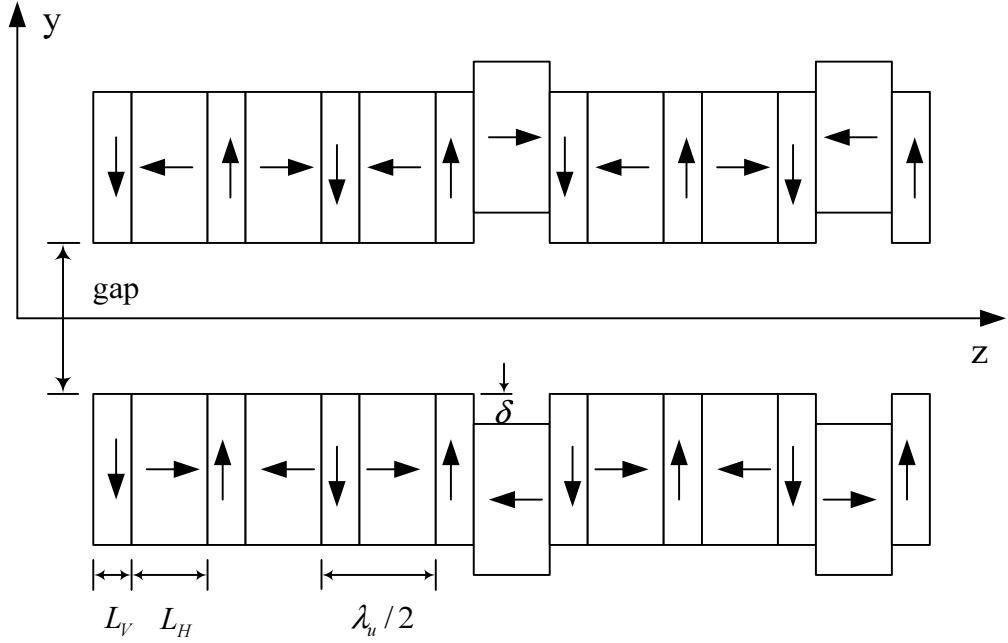

Fig.1    Scheme of modified QPU

The scheme is applied to the undulator of HLS with the electron beam of $800\,MeV$, $300\,mA$. To generate fundamental radiation around $5eV$, $\lambda_u = 88mm$ is selected. $\delta = 5mm$ and $\eta = \sqrt{5}$ are the same as ESRF device. The spectrum under different $L_V/L_H$ is evaluated with RADIA and SPECTRA as Tab.1 shows. [9-10]

It is seen that high harmonics till 7th are best depressed at $L_V/L_H = 14:30$ with total flux 40.9% of the fundamental. The 3rd harmonic is well reduced provided $L_V/L_H$ is sufficiently small. Note that $L_V/L_H$ with better 3rd elimination has a little drawback on other harmonics hence the whole harmonics has low sensitivity to $L_V/L_H$. The mechanism is that dimensions optimization doesn't introduce new irrational periodicity to QPU. It takes effect by modulating harmonic field, i.e., enhancing or depressing different harmonics field. Even though, it is much significant to eliminate the major contamination such as 3rd, 5th and 7th harmonic.

The optimized parameters of QPU are listed in Tab.2.

Tab.1 Optimization of $L_V/L_H$ under Gap=32mm.

| $L_V/L_H$ | Fundamental Photon[eV] | Fundamental flux density [ph/s/mrad²/0.1%] | Normalized harmonics density to fundamental | | | |
|---|---|---|---|---|---|---|
| | | | 3rd | 5th | 7th | total |
| 16:28 | 4.93 | 7.1e14 | 0.197 | 0.147 | 0.202 | 0.547 |
| 14:30 | 5.12 | 6.8e14 | 0.087 | 0.204 | 0.118 | 0.409 |
| 12:32 | 5.37 | 6.3e14 | 0.028 | 0.285 | 0.102 | 0.415 |
| 10:34 | 5.71 | 5.8e14 | 0.015 | 0.386 | 0.085 | 0.487 |
| 8:36 | 6.15 | 6.2e14 | 0.012 | 0.300 | 0.134 | 0.446 |

Tab.2 Parameters of HLS QPU

| Type | planar pure permanent magnet |
|---|---|
| Period length | 88mm |
| Period number | 19 |
| Total length | 1672mm |
| Magnet material | NdFeB |
| Remanence magnetic field | 1.26T |
| H-block size | 84*60*30mm |
| V-block size | 84*60*14mm |
| Working gap | 32-55mm |
| inter-lattice ratio η | $\sqrt{5}$ |
| $\delta$ | 5mm |

**3. Manufacturing and Measuring**

The complete device sizing 2.4×1.2×1.7 m³ is as Fig.2 shows. The mechanical system consists of magnetic structures, girders and C-frame. The retracted magnet shows a concave surface. Movement of the girders is driven by servo motors through PLC control system.

Magnetic measurements were made using the 8 m bench at Institute of High Energy Physics (IHEP). The vertical $B_y$ field is measured using a Hall probe with a sensitive area of about 1 mm in diameter. The residual error of the calibration was about $2\times10^{-5}T$. The horizontal field $B_x$ was measured using a sensor coil of dimensions 3.95×13.55×7.0 mm³ with 7000 turns and a calibrated winding area of about 0.27 m². Typical intervals of measurement were 0.5 mm and equidistant data point was taken. Field integrals were calculated by numerical integration of these field scans [11]. (Fig.3)

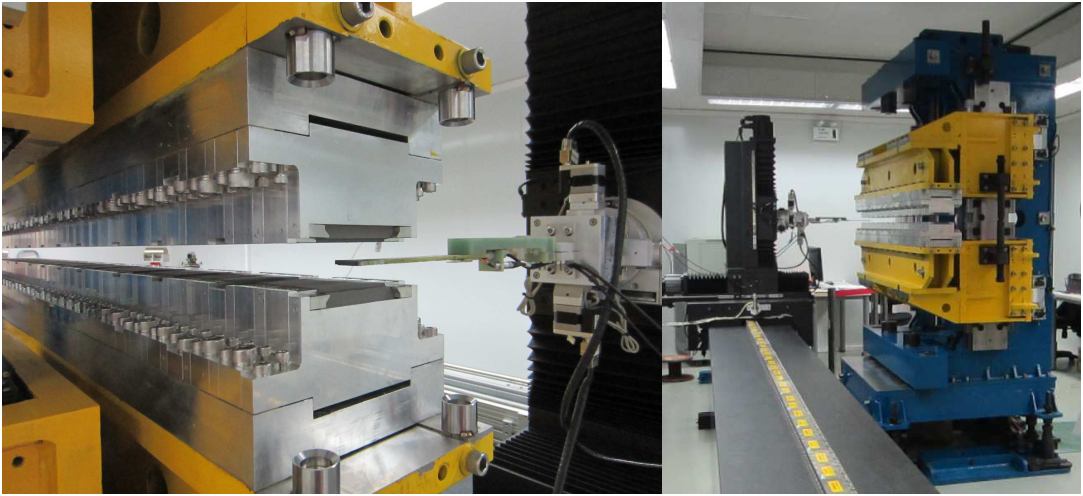

Fig.2 HLS QPU in measurement

## 4. Commissioning and Test

Field quality is urged in a region $|x|<14mm$ under tunable gaps. As irrational harmonics is mixed, the ideal phase of QPU is non-zero. Magnetic precision is guaranteed by comparing measured 1st field integrals in each period with the standard from RADIA simulations. The full specifications of the QPU including multiple integrals within tunable gaps are listed in Tab.3.

*Tab.3 Specifications of HLS QPU*

| | |
|---|---|
| 1st field integral $I_1(x)$ | <±100Gauss*cm |
| 2nd field integral $I_2(x)$ | <±20000Gauss*cm$^2$ |
| Normal and skew quadrupole integral | <50Gauss |
| Normal and skew sextupole integral | <100Gauss/cm |
| Normal and skew octupole integral | <100Gauss/cm$^2$ |

Field errors were controlled in different ways: For assembly magnets were sorted using simulated annealing. This provides some basic but not sufficient field quality. For ultimate error compensation pole height tuning and shimming was applied. Pole height tuning is responsible for on-axis $I_{1y}(x=0)$ corrections. All the other signature corrections including multiple integrals and gap-dependence kick is finished by shimming.

With appropriate symmetry applied, $I_1(x)$ and $I_2(x)$ were treated by shimming respectively with no mutual effect. Measured multiple integrals under tunable gaps is as Fig.3 shows. The potential of shimming is demonstrated.

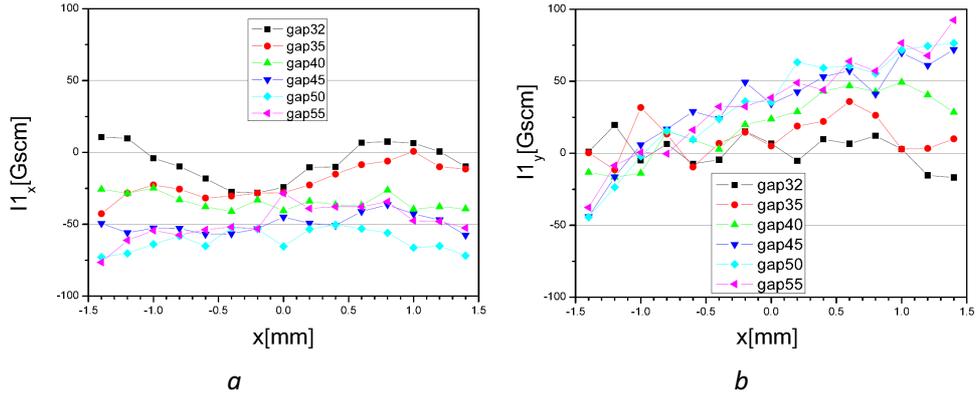

Fig.3 Measured f a) $I_{1x}(x)$ and b) $I_{1y}(x)$ under tunable gaps.

Spectrum characteristic is deduced from the measured field by SPECTRA as Fig.4 and Tab.4 shows. There is a little difference of the fundamental energy and flux between theory (design) and the deduced. It is resulted by 3% higher practical magnetic field than the design. Rational harmonics are excellently depressed as theory. By taking into account the inherent difference of each harmonics, the major contamination 3rd harmonic is depressed by 58.5 times at $Gap=32mm$. By contrast, the same figure of a ESRF QPU with uniform magnet sizes is 8.3 .[4] The deduced flux with respect to photon energy by SPECTRA[11] is plotted in Fig.5.

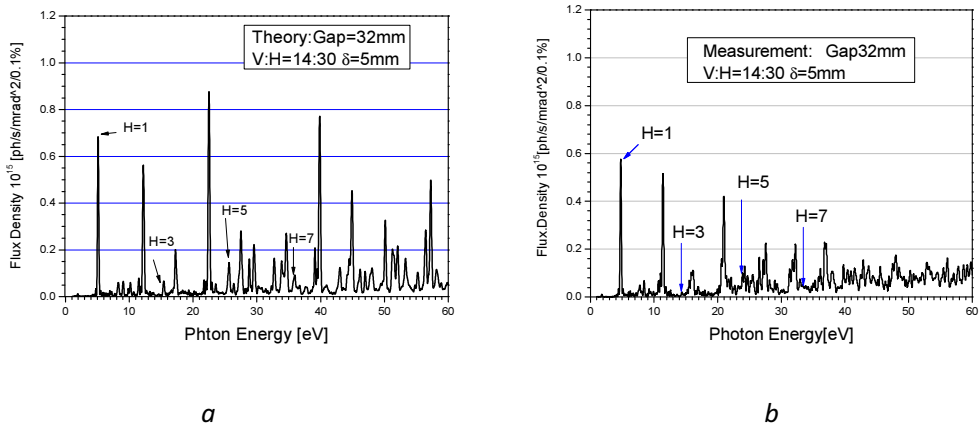

Fig.4 a) Theoretical and b) deduced radiation spectral under Gap 32mm

Tab..4    Theory and Deduced spectrum under different gaps

|  | Fundam-ental Photon[eV] | Fundamental density[ph/s/mrad$^2$/0.1%] | Normalized harmonics density to fundamental | | | |
|---|---|---|---|---|---|---|
|  |  |  | 3rd | 5th | 7th | 3rd+5th+ 7th |

| Theory Gap32 | 5.12 | 6.8e14 | 0.087 | 0.204 | 0.118 | 0.409 |
| Deduced Gap32 | 4.77 | 5.8E+14 | 0.026 | 0.151 | 0.080 | 0.256 |
| Theory Gap55 | 20.71 | 1.7e15 | 0.022 | 0.161 | 0.060 | 0.243 |
| Deduced Gap55 | 19.52 | 1.4E+15 | 0.04 | 0.118 | 0.106 | 0.272 |

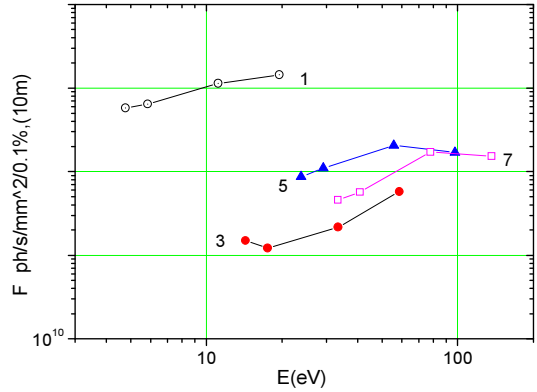

Fig. 5    Flux of HLS QPU [13]

**4. Summary**

A practical modified QPU is ideveloped for HLS. The Field tests support that harmonics emission is much optimized. Depression ratio of 3$^{rd}$ harmonics is increased by an order of magnitude than current QPU. The significance of the magnet dimensions optimization to harmonics elimination is verified. Magnetic performance is tested in a good field region $|x|<14mm$ under $Gap=32-55mm$. Multiple 1$^{st}$ integrals, on-axis 2$^{nd}$ integrals overcoming gap-dependence kicks are finished by shimming. The potential of shimming is demonstrated.

**5. Acknowledgements**

The work is supported by development fund of Institute of High Energy Physics, Chinese academy of Sciences. Thank Hefei Light Source for their helpful discussions.